# A Solution to Kolmogorov-Feller Equation and Pricing of Option

Prof. Dr. *Kim Ju Gyong* , *Choe Il Su*

College of Mathematics, **Kim Il Sung** University

**Abstract** We study the solution to Kolmogorov-Feller equation and by using it develop pricing formulas of well known some option with jump.

The great leader comrade **Kim Jong Il** said as follows.

"**In addition, we must intensify the study of mathematics, physics, biology and other basic science so that they contribute positively to the national economy and the development of science and technology.**"("**Kim Jong Il** Selected Works" Vol. 12, 202p)

In this paper we is to develop pricing formulas for several well known options using a solution to Kolmogorov-Feller equation by generating operator of finite-type Levy random process.

In finance mathematics many researchers considered the pricing of option describing by Levy processes with jump [1－10]. In [4] and [6] the pricing formula of option is fully developed. L. S. Jiang(2003)[4] derived the pricing formula of European option by numeric solution in the case of pure Poisson process. And S.Jaimungal(2006)[6] derived the pricing formula of a calamity option as integral type in the case of compound Poisson process. The other papers discuss the case of Levy processes with jump but those did not fully develop the formula and derived only the equation that the precing process satisfies or its numerical solution [10]. To drive the pricing formula of option in the case of a Brownian motion must be solved the Kolmogorov equation and in the case of Levy processes with jump must be solved Kolmogorov-Feller equation. The Kolmogorov equation is well known as heat equation and solution is also developed but Kolmogorov-Feller equation is a problem of second partial differential- integral equation and its solution is limited to some special problem [11].

## 1. A solution method to Kolmogorov-Feller equation

Let's see the following Kolmogorov-Feller equation

$$\left. \begin{array}{l} \left(\dfrac{\partial}{\partial s} + L_s\right)u(s,\ x) = 0,\quad (s,\ x) \in [0,\ T] \times \boldsymbol{R}^n \\ u(T,\ x) = \varphi(x) \end{array} \right\} \quad (1)$$

, where operator $L_S$ is

$$(L_s f)(x) = \sum_i a_i(s)\frac{\partial}{\partial x_i}f(x) + \frac{1}{2}\sum_i\sum_j a_{ij}(s)\frac{\partial^2 f(x)}{\partial x_i \partial x_j} + \int_{|z|>0}[f(x+c(s,\ z)) - f(x) - \sum_i c_i(s,\ z)\frac{\partial}{\partial x_i}]\nu(dz)$$

and $\nu(dz)$ is the Levy measure.

**Assumption 1** (a) the functions $a_{ij}(t)$, $i$ $(j = 1, 2, \cdots, n)$, $a_i(t)$ $(i = 1, \cdots n)$ are of $c^\infty$ – class.

(b) $c_i(t, z)$, $i = \overline{1, n}$ is $c^\infty$ – function by $t$ and $\int_0^T \int_{R^n} |c_i(t, z)| \nu(dz) dt < +\infty$, $i = \overline{1, n}$.

For any $\theta \in R^n$ ($\theta \neq 0$), there exists $\lambda > 0$ such that $\sum\sum a_{ij}(t)\theta_i\theta_j \geq \lambda \|\theta\|^2$.

Let decompose operator $L_s$ into $A_s$ and $B_s$ as following

$$\left. \begin{aligned} (A_s f)(x) &= \sum_i a_i(s) \frac{\partial}{\partial x_i} f(x) + \sum_i \sum_j a_{ij}(s) \frac{\partial^2 f(x)}{\partial x_i \partial x_j} \\ (B_s f)(x) &= \int_{|z|>0} \left[ f(x + c(s, z)) - f(x) - \sum_i^n c_i(s, z) \frac{\partial}{\partial x_i} f(x) \right] \nu(z) \end{aligned} \right\} \quad (2)$$

$$\left. \begin{aligned} \left(\frac{\partial}{\partial s} + A_s\right) u(s, x) &= 0 \\ u(T, x) &= \varphi(x) \end{aligned} \right\} \quad (3)$$

$$\left. \begin{aligned} \left(\frac{\partial}{\partial s} + B_s\right) u(s, x) &= 0 \\ u(T, x) &= \delta(x) \end{aligned} \right\} \quad (4)$$

**Theorem 1** If (a), (c) of Assumption 1 is satisfied, then the solution of equation (3) is

$$u(s, x) = (2\pi)^{-n} \int_{R^n} e^{-i(\theta, x)} \hat{\varphi}(\theta) \exp\left\{ -\frac{1}{2} \int_s^T \theta' A(t) \theta dt - i \int_s^T (a(t), \theta) dt \right\} d\theta$$

, where $\hat{\varphi}$ is Fourier transform of $\varphi$ and $A(t) = (a_{ij}(t))$, $a(t) = (a_i(t))$.

**Theorem 2** If (b) of Assumption 1 is satisfied, then the solution of equation (4) is

$$u(s, x) = (2\pi)^{-n} \int_{R^n} \exp\left\{ -i(\theta, x) + \int_s^T \int_{R_0^m} [\exp(-i(\theta, c(\tau, z))) - 1 - (\theta, c(\tau, z))] \nu(dz) d\tau \right\} d\theta$$

**Theorem 3** Suppose $u_1$, $u_2$ are a solution of (3), (4). Then $u = u_1 * u_2$ is a solution of (1), where * is convolution with respect to $x$.

## 2. Transition probability density function of several Levy processes.

**Theorem 4** Suppose $P(x)$ is the generalized density function of random variables $\xi_i$ then the transition probability density function of compound Possion process $\{X_t\}_{t \in [0, +\infty)}$ is

$$f(s, x; t, y) = \sum_{n=0}^{\infty} \frac{[\lambda(t-s)]^n}{n!} e^{-\lambda(t-s)} P^{*n}(y - x), \quad (5)$$

where $P^{*0}(y - x) = \delta(y - x)$, $P^{*n}$ is n-multiple convolution of generalized density functions.

The transition probability density function of Levy processes $\xi = \{\xi_t(\omega)\}_{t \in [0, +\infty)}$ defined as $\xi_t(\omega) = \gamma t + \sigma W_t + \sum_{i=1}^{N(t)} \xi_i$ is

$$\sum_{n=0}^{\infty}\sum_{k=0}^{n \cdot m}\frac{[\lambda(t-s)]^n}{n!}e^{-\lambda(t-s)}C_{n \cdot m}^k p^k q^{nm-k}\frac{1}{\sqrt{2\pi(t-s)\delta^2}}\exp\left\{-\frac{[y-x-\gamma(t-s)-k]^2}{2\delta^2(t-s)}\right\},$$

where the distribution of $\xi_i$ is $B(m, p)$.

And $P_l(\cdot)$ is generalized density function of $l$-th distributions.(i.e. I-unit, U-general discrete, G-geometry, B-binomial, $P_0$-Possion, E-exponential, N-Normal distributions)

### 3. Pricing of several options

In finance market there exist two assets like following.
$$dB_t = rB_t dt, \quad B_0 = 1 \tag{6}$$

$$dS_t = S_t\left[b(t)dt + \sigma(t)dW_t + \int_{|z|>0}c(z)\mu(dt, dz)\right], \tag{7}$$
$$S_0 = S, \quad t \in [0, T]$$

where r is the interest rate, $b(t)$ is the process of "appreciation rate", $\sigma(t)$ is the processes of "volatilities" of the size of continuous risk, $c(z)$ is the size of jump risk.

**Assumption 2** (a). $b(t), \sigma(t)$ is measurable function bounded on $[0, T]$.

(b) $\int_{|z|>0}|c(z)|\nu(dz) < +\infty$

(c) $\mu(dt, dz)$ is the integer random measure consisted of Compound Possion processes, its compensator $\nu(dz) = \lambda p(z)dz$ and $p(z)$ is the generalized density function of $\xi_n$.

If (b) of Assumption 2 is satisfied then equation (7) in the standard complete market is expressed as
$$dS_t = S_t\left\{rdt + \sigma(t)d\widetilde{W}_t + \int_{|z|>0}c(z)(\widetilde{\mu}-\widetilde{\nu})(dt, dz)\right\}, \quad S_0 = S, \quad t \in [0, T] \tag{8}$$

Then Black-Scholes operator is given by
$$LV = \frac{\partial V}{\partial t} + \frac{\sigma^2(t)}{2}S^2\frac{\partial^2 V}{\partial S^2} + rS\frac{\partial V}{\partial S} - rV + \int_{R_0}\left[V(t, S+SC(z)) - V(t, S) - SC(z)\frac{\partial V}{\partial S}\right]\widetilde{\nu}(dz) \tag{9}$$

And the pricing formula of option can be derived from the solution of the equation
$$LV = 0, \quad V(T, s) = \varphi(s) \tag{10}$$

**Theorem 5** The fundamental solution of equation (10) corresponding operator $L$ (form (9)) and Levy measure $\widetilde{\nu}(dz) = \widetilde{\lambda}p_l(z)dz$ is $f(s, x; t, y) = e^{-\gamma(t-s)}J_l(s, x; t, y)$.

**Corollary** The solution of equation (12) is $V(t, s) = \int_{-\infty}^{+\infty}\varphi(y)f(t, s; T, y)dy$.

From the corollary the earlier pricing formulas derived by geometry Brownian motion can be generalized into pricing formula of option derived by geometry Levy processes (Consist compound Possion process). Let's see the cases of some options.

**Option by terminal problem** European option, exchange option, exotic exchange option and cash-or-nothing option are all determined its payoff amount at the maturity of contract. Therefore they can be handled by terminal problem of partial differential-integral equation. If the terminal condition of equation (10) is $V(T, S) = \varphi(S) = (S - K)^+$, then from the Corollary of Theorem 5 the price of a European call option is given by formula

$$S \sum_{|l(m)|=0}^{\infty} \frac{[\tilde{\lambda}(1+c(a))]^{l(m)}(T-t)^{|l(m)|}}{l(m)!} e^{-|\tilde{\lambda}(1+c(a))|(T-t)} \Phi(d_+^s) - Ke^{-\gamma(T-t)} \sum_{|l(m)|=0}^{\infty} \frac{\tilde{\lambda}^{l(m)}(T-t)^{|l(m)|}}{l(m)!} e^{-|\tilde{\lambda}|(T-t)} \Phi(d_-^s)$$

, where $d_\pm^l = \left[\ln S/k + (\gamma - \bar{c}_l)(T-t) \pm \sigma^2(t, T)/2 + \sum_l \right] / \sqrt{\sigma^2(t, T)}$.

According to the Levy measure $\nu(dz) = \lambda p_l(z)dz$ $(l = i, s, d, b, g, p, N)$, the pricing of a European call option is easily derived.

By the same method, the pricing of other option (exchange option, exotic exchange option and cash-or-nothing option) can be derived.

**Option as a boundary problem** There exist not only the options that have effect at maturity of contract, but the options that effect before contract period under certain condition. Barrier option, time dependent barrier option, path dependent option and so forth are of this type.

Let's discuss the down-and out option among these.

The pricing of this option is boundary problem of partial differential-integral equation

$$\left. \begin{array}{l} LV(t, S) = 0, \quad 0 \leq t \leq T, B < S \\ V(T, S) = (S - K)^+ \\ V(t, B) = 0, \quad 0 < t < T \end{array} \right\}.$$

The pricing of down-and-out call option is

$$V(t, S) = S \cdot F_{jl}(d_+^l j(1+c(\cdot))) - ke^{-\gamma(T-t)} F_{Jl}(d^l - j) - S \cdot \left(\frac{B}{S}\right)^{\frac{2}{\sigma^2}t+1} F_{Jl}(\mu_+^l j(1+c(\cdot))] - ke^{+(T-t)} F_{Jl}(\mu_-^l),$$

where $d_\pm^l$ is the same as in the table 3 and $\mu_\pm^l = \dfrac{\ln\dfrac{B^2}{Sk} + (\gamma - \bar{c}_l)(T-t) \pm \dfrac{\sigma^2(T-t)}{2} + \sum_l}{\sigma\sqrt{(T-t)}}$.

By the same method, pricing formulas of down-and-out put option, up-and-out call (put) option, down-and-in option, up-and-in option can be derived.

**References**
[1] S. Asmussen; Stochastic Processes and Their Applications, **109**, 79, 2004.
[2] P. Boyle et al.; Insurance: Math. Econo., **40**, 267, 2007.
[3] J. Jang; Insurance: Math. Econo., **41**, 62, 2007.
[4] L. S. Jiang; J. of Computational and Applied Mathematics, **156**. 23, 2003.
[5] S. Emmer et al.; Finance and Stochastics, **8**, 17, 2004.
[6] S. Jaimungal et al.; Insurance: Math. Econo., **38**, 469, 2006.
[7] B. Roos; Insurance: Math. Econo., **40**, 403, 2007.
[8] G. Stoica; Proceeding of the Americal Mathematical Society, **130**, 1819, 2001.
[9] R. J. Elliott; Finance and Stochastics, **10**, 250, 2006.
[10] O. Kudryavtsev et al.; Finance and Stochastics, **13**, 531, 2009.
[11] Yong Ren; J. of Com. Appl. Math., **233**, 901, 2009.